\documentclass[twoside,leqno,twocolumn]{article}

\usepackage{ltexpprt}
\usepackage{color}
\usepackage{colortbl}
\usepackage{multicol}
\usepackage{booktabs, tabularx, makecell, multirow,caption}
\usepackage{subcaption}
\usepackage{amsmath}
\usepackage{algorithm}
\usepackage{balance}
\usepackage{thm-restate}
\usepackage[end]{algpseudocode}
\usepackage{xspace}
\usepackage[symbol]{footmisc}
\usepackage{booktabs, tabularx, makecell, multirow}
\usepackage[shortlabels]{enumitem}
\usepackage{wrapfig}
\usepackage{graphicx}
\usepackage{todonotes}
\usepackage[sort&compress,comma,square,numbers]{natbib}

\usepackage{etoolbox}
\DeclareMathOperator{\dis}{dist}
\DeclareMathOperator{\dist}{dist}

\DeclareMathOperator{\EPL}{Obj}

\DeclareMathOperator{\condentropy}{f_{\text{CE}}}
\DeclareMathOperator{\avgdeg}{\Delta_{\text{avg}}}

\newcommand{\problem}{\emph{MWASP}\xspace}

\newcommand{\algoname}[1]{\texttt{#1}}
\newcommand{\superalgorithm}{\algoname{SpiderDAN}\xspace}

\newcommand{\appsymb}{$\star$}
\newcommand{\appref}[1]{{\hyperref[proof:#1]{(\appsymb)}}}

\newcommand{\appendixproof}[2]{%
  \gappto{\appendixProofs}
  {
    \subsection{Proof of \cref{#1}}\label{proof:#1}
    #2
  }
}

\usepackage{hyperref}
\definecolor{navyblue}{rgb}{0.0, 0.0, 0.5}
\definecolor{darkspringgreen}{rgb}{0.09, 0.45, 0.27}
\hypersetup{
 colorlinks,
 citecolor = navyblue,
 urlcolor  = blue,
 linkcolor = darkspringgreen,
}

\usepackage{cleveref}

\begin{document}
\date{}


\title{SpiderDAN: Matching Augmentation in Demand-Aware Networks}
\author{Aleksander Figiel\thanks{Technische Universität Berlin, Germany}
    \and Darya Melnyk\footnotemark[1]\phantom{\footnote{This work is supported by the European Research Council (ERC) under grant agreement No.$\,$864228$\,$(AdjustNet),~2020-2025.}}
    \and André Nichterlein\footnotemark[1]
    \and Arash Pourdamghani\footnotemark[1]
    \and Stefan Schmid\footnotemark[1] }

\maketitle







\begin{abstract} \small\baselineskip=9pt 
Graph augmentation is a fundamental and well-studied problem that arises in network optimization. We consider a new variant of this model motivated by reconfigurable communication networks.
In this variant, we consider a given physical network and the measured communication demands between the nodes. Our goal is to augment the given physical network with a matching, so that the shortest path lengths in the augmented network, weighted with the demands, are minimal.
We prove that this problem is NP-hard, even if the physical network is a cycle.
We then use results from demand-aware network design to provide a constant-factor approximation algorithm for adding a matching in case that only a few nodes in the network cause almost all the communication. 
For general real-world communication patterns, we design and evaluate a series of heuristics that can deal with arbitrary graphs as the underlying network structure.
Our algorithms are validated experimentally using real-world traces (from e.g., Facebook) of data centers.
\end{abstract}

\section{Introduction}
This paper considers a network augmentation problem that is motivated by emerging data center technologies~\cite{51587,osn21,helios,rozenschiff_et_al:LIPIcs.OPODIS.2022.25}. 
The idea of these technologies is to enable demand-aware networks by using reconfigurable optical switches, on top of an existing demand-oblivious data center topology based on electrical switches.
%
An optical circuit switch allows one to directly connect (i.e., match) each data center rack to at most one other rack via an optical link, thus forming a (perfect) matching on top of an existing electrical network~\cite{10.1145/1851182.1851222,10.1145/1851182.1851223,10.1145/2716281.2836126}. This allows to reduce the number of hops traversed by communicated bit, and hence increases the bandwidth. See for example Jupiter Evolving for a recent solution deployed by Google ~\cite{51587}.



From a theoretical perspective, network augmentation has been studied under two optimization criteria so far: minimizing the diameter of the network (i.e., the worst-case communication cost)~\cite{chung1987diameters,https://doi.org/10.1002/jgt.10122,https://doi.org/10.1002/jgt.3190080408,decreasingDiameterAlon,diameterMinimization}, and minimizing average shortest path length between any two nodes~\cite{watts1998collective,10.1145/335305.335325,80gozzard2018converting,ShortcutAddition,82papagelis2011suggesting,83parotsidis2015selecting}. However, previous work has mostly assumed that the communication demand between the nodes is evenly distributed, i.e., that any two nodes are equally likely to communicate, independent of their role in the network.
We drop this assumption in this work and arrive at the following graph problem:
Given a graph (the existing infrastructure network) and a demand matrix (encoding the communication demand between all pairs of nodes), 
our goal is to compute a matching to add so that the weighted average path length in the augmented network is minimized where the weights are given by the demand matrix.
We call this problem \textsc{Minimizing Weighted Average Shortest Path Length via Matching Addition}~(\problem), see Section~\ref{sec: model} for a formal definition. 
This approach can be used to reconfigure the network after some time to adjust to the new demand. 




\subparagraph*{Our Contribution.} 


We analyze the complexity and approximability of \problem.
We start by showing that it is NP-hard, already if the underlying infrastructure graph is a ring and each node communicates to at most two other nodes (i.\,e., the demand matrix is extremely sparse). 

We further propose a constant-factor approximation algorithm for \problem on connected underlying infrastructure graphs of constant maximum degree. 
This algorithm groups small segments of nodes that are in close proximity into super-nodes and then connects these super-nodes with a known construction from demand aware networks~\cite{avin2017demand}.
The idea behind the grouping is that nodes can help their neighbors of high demand connect to other high-demand nodes.
This construction limits us to highly skewed demand matrices where only a few nodes in the network cause almost all the communication.

We consider synthetically generated communication demands, as well as real-world datasets; in particular a dataset collected from Facebook~\cite{facebook}.
We further test our algorithms on various infrastructure graphs, including symmetric structures like rings, 2D and the 3D torus, as they are often used in distributed architectures~\cite{1992xix}.
Our heuristics and approximation prove highly efficient and effective on these datasets:
They display good approximation factors on the small synthetic instances where we do know the optimum.
Moreover, the approximation and 
our heuristics scale well on the real-world datasets.


\subparagraph*{Related Work.}

Graph augmentation has been widely studied from the perspective of optimizing communication in a peer-to-peer network.
One goal is to minimize the diameter of a graph~\cite{chung1987diameters}, i.e., to improve the worst-case cost of routing. 
\citet{https://doi.org/10.1002/jgt.3190110315} showed that the problem of minimizing the number of edges that one needs to add to a graph in order to reduce its diameter to $d$ is NP-complete. Following this result, lower and upper bounds on the number of additional edges have been derived for  cycles~\cite{https://doi.org/10.1002/jgt.10122,https://doi.org/10.1002/jgt.3190080408,decreasingDiameterAlon}, paths~\cite{https://doi.org/10.1002/jgt.3190080408}, $(t+1)$-edge connected graphs~\cite{https://doi.org/10.1002/jgt.3190080408}, and graphs of bounded maximum degree~\cite{https://doi.org/10.1002/jgt.3190080408,decreasingDiameterAlon}. In another variant, the goal is to add a fixed number of edges such that the diameter is minimized~\cite{10.1145/335305.335325}. \citet{diameterMinimization} generalize this setting to adding up to $\delta$ edges per node, which corresponds to adding a matching for $\delta=1$. Both variants have been shown to be NP-hard and algorithms with logarithmic approximation ratios (in the number of added edges) have been proposed in the respective papers.
This problem has also been considered from the viewpoint of opinion polarization in social networks. Here, the goal is to reduce polarization, i.e., the distance between pairs of nodes located in different groups, by adding a small number of edges to the graph~\cite{https://doi.org/10.1111/itor.12854,10.1145/3437963.3441825,10.1145/3018661.3018703}.

Graph augmentation has also been studied in the context of small-world networks. In this setting, the main question is how many edges one needs to add to a graph to minimize the average shortest path length. \citet{watts1998collective} proposed a first model where the nodes are placed on a circle, and edges representing ``long-range'' contacts are added randomly to each node.
Kleinberg~\cite{10.1145/335305.335325} introduced a formal model where the nodes are placed on a grid, and each node is allowed to add one shortcut edge following the inverse power distribution. Kleinberg showed that such a small-world network has an average path length that is logarithmic in the number of nodes.
\citet{ShortcutAddition} introduce edge weights in their model and allow the addition of a constant number of shortcut edges to the graph. They show that this problem is NP-complete and provide constant approximation algorithms for the weighted and the unweighted cases. 
Since many of the problem variants have been proven to be hard, different heuristics for adding a small number of edges to a network have been considered in the literature~\cite{82papagelis2011suggesting,83parotsidis2015selecting,80gozzard2018converting}. 


~
In this paper, we look at a demand-aware average shortest path length as the quality measure of our generated network. 
Our motivation to study an augmentation variant with a matching comes from demand-aware data center networks~\cite{51587,osn21}: emerging optical switching technologies make it possible to enhance a given network with a demand-aware matching. 
This model has been considered before also in the theoretical literature: \citet{spaa21rdcn} consider a setting where packets need to be routed from sources to destinations via a matching. The authors present a stable matching that is updated over time in an online fashion. \citet{hanauer2023dynamic} presents a dynamic setting in which a weighted $k$-disjoint matching is recomputed dynamically based on changing demand. 
Other algorithms aim at minimizing the load or the congestion of routing in demand-aware networks~\cite{10.1145/3597200,infocom24dan}. 

Also, demand-aware network design has been considered in the literature, however, not with a single matching. \citet{avin2017demand}
propose designing demand-aware networks from scratch by building a bounded degree graph. They provide a lower bound on the average path length based on the entropy of the demand matrix, and present a constant approximation algorithm for sparse demand graphs. \citet{hanauer2022fast} study algorithms to find $k$ disjoint heavy matchings in a graph. They show that this problem is NP-hard, and propose different approximation algorithms for the problem.


\section{Model}
\label{sec: model}

\noindent \textbf{Infrastructure graph.} We are given a set of nodes $V=\{v_1, \dots, v_n\}$ that communicate over an underlying \emph{infrastructure graph} $G$, this graph corresponds to the physical network. 
The infrastructure graph is assumed to have a non-constant diameter and a large number of nodes $n$, where $n$ is even (so that there always exists a perfect matching).

\noindent \textbf{Demand matrix.}
The communication pattern is described by an $n \times n$ \emph{demand matrix} $D$. In this matrix, $D_{u,v}$ indicates the probability with which a node $u$ communicates to a node $v$. Observe that the demand matrix is normalized, i.e., $\sum_{u,v\in V, u\neq v} D_{u,v} = 1$, and the demand from one node to itself is $0$, i.e., $D_{v,v} = 0\ \forall v\in V$. 
For simplicity, we assume that the demand matrix is symmetric. 
Thus, the demand matrix encodes an edge-weighted, simple, undirected graph (zero-weight edges are omitted).
We use $\dis_G(u,v)$ to denote the \emph{distance} (the length of a shortest path between) $u$ and~$v$ in~$G$. 
Our objective is to minimize the weighted average shortest path length $\EPL_D(G) = \sum_{u,v\in V} D_{u,v}\cdot \dis_G(u,v)$ in $G$ for the given demand matrix $D$. 

\noindent \textbf{Optimization objective.} Our goal is to add a perfect matching $M$ to the set of edges of~$G$ that minimizes the weighted average shortest path length in this augmented graph~$G+M$. 
We consider the case where the added matching edges behave the same as the edges of the underlying graph, i.e., they are indistinguishable from the edges of the infrastructure graph in terms of their weight and length. 
We now define \textsc{Minimizing Weighted Average Shortest Path Length via Matching Addition (\problem)} as finding for a given graph~$G$ a matching $M$ that minimizes~$\EPL_D(G + M)$.





\section{NP-Hardness}
Before discussing the approximation algorithm in the next section, we state that even restricted to very simplistic underlying infrastructure graphs and sparse demand matrices, the problem remains NP-hard.
Thus, there is (probably) no polynomial-time algorithm computing optimal solutions for more interesting real-world infrastructure graphs and demand matrices.

\begin{theorem}
\label{thm:np-hardness}
    \problem is NP-hard, even if the underlying graph is a cycle and every row and column of~$D$ has at most two non-zero elements.
\end{theorem}

\begin{proof}
	\newcommand{\lone}{\ensuremath{\alpha}}
	\newcommand{\lmin}{\lone}
	\newcommand{\ltwo}{\ensuremath{\beta}}
	\newcommand{\lmax}{\ltwo}
	We reduce from \textsc{Vertex Cover}, which remains NP-hard on graphs of maximum degree three~\cite{GJ79}.
	Given a graph~$G=(V,E)$ of maximum degree three and an integer~$k$, the question is whether there is a vertex cover of size at most~$k$, that is, whether there is a vertex subset~$S \subseteq V$, $|S| \le k$, such that for each~$e\in E$ we have~$e \cap S \neq \emptyset$?
	Considering such a \textsc{Vertex Cover} instance~$(G,k)$, we build an instance~$(G' = (V',E'),D,b)$ of the decision version of \problem where~$G'$ is a cycle and~$b$ is the cost bound, that is, the question is if there is a matching~$M$ so that~$\EPL_D(G' + M)\le b$?
	
	To better distinguish the graphs~$G$ and~$G'$ we use the term \emph{vertices} for~$G$ and \emph{nodes} for~$G'$.
	Throughout our construction we want to enforce certain edges to be in a solution; denote with~$q$ the number of these edges.
	For each of these edges, we set the demand in~$D$ to a high number~$\lmax$ and set the cost bound~$b$ to satisfy~$q \lmax < b < (q+1)\lmax$.
	(For ease of presentation, we do not normalize~$D$. By dividing~$b$ and each entry in~$D$ by the sum of entries in~$D$ we could normalize~$D$ without changing the problem).
	Thus, the budget constraint enforces that the endpoints of each of the~$q$ edges need to have a distance of one in the resulting graph.

	Subsequently, we describe our gadgets, each of which forms a path of nodes in~$G'$. 
	In the end, these paths will be connected to form the ring~$G'$.
    Refer to Figure~\ref{fig: hardness} for a sketch of the gadgets and their interactions.
    
\begin{figure}[t]
	\centering
	\includegraphics[width = 0.9\linewidth, clip]{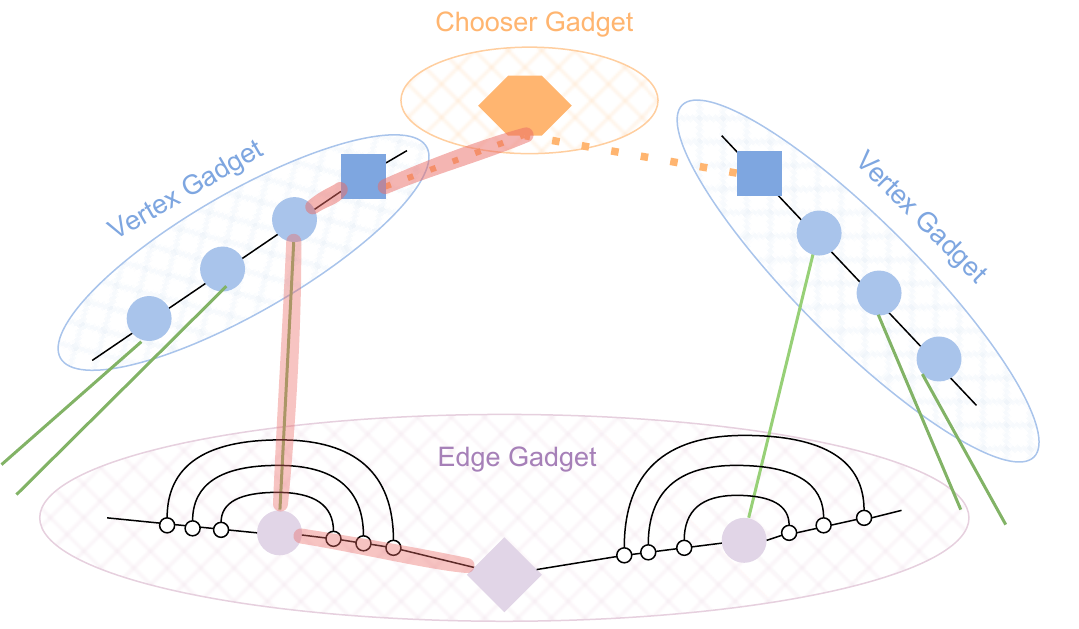}
	\caption{
		A schematic picture of our construction. 
		The green and solid green edges incident to the vertex gadgets and the black edges within the edge gadget are forced to be in the matching.
		One vertex in the chooser gadget has a demand to the middle diamond vertex in the edge gadget in the bottom but cannot reach it directly.
		The only two options, indicated by the dotted edges: 
		Connect a vertex in the chooser gadget to the helper vertex~$h_v$ (denoted by the squares) in one of the two vertex gadgets that represent the endpoints of the edge.
		In the example, the left option is chosen and the path to the diamond vertex in the bottom goes via the vertex gadget into the edge gadget.
		Note that the long path between the left circle node~$e^\ell$ and the diamond node~$e^\text{mid}$ in the edge gadget prevents traversing the edge gadget from one circle node ($e^\ell$) to the other ($e^r$).
	} 
	\label{fig: hardness}
\end{figure}
 
	\noindent \textbf{Vertex gadget.} 
	Consider a vertex $v \in V$ incident to between one and three edges (vertices of degree zero can be ignored). 
	Add a path on $\deg(v)+1$ nodes to $G'$.
	These $\deg(v)+1$ nodes are a \emph{helper} node $h_v$ (first position in the path) and \emph{vertex} nodes $v^1, \dots, v^{\deg(v)}$.
	Each vertex node~$v^i$ corresponds to one edge incident to~$v$.

	\noindent \textbf{Edge gadget.} 
	For each edge $e \in E$, add a path~$P^e$ on~$4\cdot\lone-1$ nodes~$e_1,\ldots,e_{4\lone - 1}$ to~$G'$ ($\lone$ will be specified later).
	The unique \emph{middle} node in this path is~$e^\text{mid} = e_{2\lone} $ and the nodes at distance exactly~$\lone$ from~$e^\text{mid}$ are the \emph{edge} nodes $e^\ell = e_{\lone}$ and~$e^r = e_{3\lone}$ (for left and right).
	The remaining~$4(\lone-1)$ nodes are \emph{dummy} nodes whose purpose is to ensure the distances between~$e^\text{mid}$ and~$e^\ell,e^r$.
	To ensure the matching cannot ``disrupt'' the distances on the path, we set for each~$i\in [\lone-1]$ the demand of the pairs~$D(e_{\lone-i},e_{\lone+i}) = D(e_{3\lone-i}, e_{3\lone+i}) = \lmax$.
	Add a dummy node~$e_0$ at the beginning of the path~$P^e$ and set a demand~$D(e^\text{mid},e_0) = \lmax$. 
	Let~$u,v$ be the endpoints of~$e$, that is, $e = \{u,v\}$.
	Add the demand~$D(e^\ell,v^i) = D(e^r,u^j) = \lmax$ where~$v^i,u^j$ are the vertex nodes corresponding to~$e$.
	Here the mapping of~$e^\ell$ to~$v^i$ and~$e^r$ to~$u^j$ is arbitrarily chosen, but fixed.

	\noindent \textbf{Chooser gadget.}
	Add~$m = |E|$ \emph{demand} nodes~$d_1,\ldots,d_m$, connected on a path.
	For an edge~$e_i \in E$ the demand node~$d_i$ has a demand of~$1$ to~$e_i^\text{mid}$, that is, $D(d_i, e_i^\text{mid}) = 1$.
	For all but~$k$ of these demand nodes we force edges in the matching as follows.
	To this end, assume without of generality that~$k$ and~$m$ have the same parity (otherwise double~$G$ and~$k$).
	Thus~$m-k$ is even.
	For~$i \in [(m-k)/2]$ add the demand~$D(d_i,d_{m+1-i}) = \lmax$.
	The remaining~$k$ demand nodes are free to be connected to the vertex nodes (the idea is that they select a vertex cover).
	
	\noindent \textbf{Gap gadget.}
	To ensure that different gadgets are far apart on the ring, we add a \emph{gap paths}.
	Each gap path consists of~$4\lmax$ \emph{gap} nodes~$g_1, \ldots, g_{4\lmax}$.
	For each~$i \in [\lmax]$ the demands are~$D(g_{4i-3},g_{4i-1}) = D(g_{4i-2},g_{4i}) = \lmax$.

	\noindent \textbf{The overall construction.} 
	Put all vertex, edge gadgets and the chooser gadget on the ring (in arbitrary order), ensuring that between any two of these gadgets a gap gadget is placed.
	Set all the demands that were not mentioned to zero.
	Note that~$b < (q+1)\lmax$ and all numbers are polynomially bounded in the input.
	\noindent \textbf{Correctness.}
	We show that~$(G,k)$ is a yes-instance of \textsc{Vertex Cover} if and only if~$(G',D,b)$ is a yes-instance of \problem.
	
	``$\Rightarrow$'': Let~$S \subseteq V$ be a vertex cover of size at most~$k$ for~$G$.
	Denote~$S = \{s_1,\ldots, s_k\} \subseteq V$.
	Then add to~$G'$ the following matching~$M$: 
	For each pair~$\{u,v\}$ with demand~$D(u,v) = \lmax$ add~$\{u,v\}$ to~$M$.
	Moreover, for each vertex~$s_i \in S$ add the edge~$\{d_{(m-k)/2 + i},h_{s_i}\}$.
	We claim that this matching incurs a cost of at most~$b$.
	By construction, each pair~$\{u,v\}$ with demand~$D(u,v) = \lmax$ has distance one in~$G' \cup M$.
	Thus, these pairs contribute~$q\lmax$ to the cost.
	The only other non-zero demands are between the demand nodes and the middle nodes in the edge gadgets.
	Consider demand node~$d$ having demand one to~$e^\text{mid}$.
	Since $S$ is a vertex cover, there is an~$s \in S$ with~$e \cap s \ne \emptyset$.
	Hence, the edge~$\{d',s\}$ is in~$M$ for some demand node~$d'$.
	Moreover, $M$ also contains the edge~$\{s^i, e^\ell\}$ or the edge~$\{s^i, e^r\}$ for some $i \in [3]$.
	Thus, in~$G'\cup M$ there is the path~$P$ from~$d$ to~$e^\text{mid}$ via~$d'$, $h_{s}$, $s^i$, and either~$e^r$ or~$e^\ell$.
	Every demand node has a distance less than~$m$ to each other demand node in~$G'$. 
	Moreover, $e^r$ and~$e^\ell$ have distance $\lone$ to~$e^\text{mid}$.
	Hence, $P$ has distance at most~$m + \lone + 3$.
	This gives a cost of at most~$m(m + \lone + 3)$ for the demands of value one.
	Thus, the overall cost is at most~$b$.
	
	``$\Leftarrow$'':
	Let~$M$ be a matching added to~$G'$ such that the cost is at most~$b$.
	We claim that in~$G$ there is a vertex cover~$S \subseteq V$, $|S| \le k$, formed by the vertices whose helper nodes are matched to demand nodes in~$M$, formally, $S := \{v \in V \mid \{d_i,h_v\}\in M \land i \in [m]\}$.
	
	Denote with~$M'$ the pairs~$\{u,v\}$ with demand~$D(u,v) = \lmax$.
	Since~$b < (q+1)\lmax$, it follows that~$M' \subseteq M$.
	The only nodes not matched in~$M'$ are~$k$ demand nodes~$d_i$, $i \in \{(m-k)/2+1, \ldots, m - (m-k)/2\}$, and all helper nodes.
	Thus, $|S| \le k$.
	It remains to show that~$S$ is indeed a vertex cover.
	
	The remaining budget for the demands between the demand nodes and the middle nodes is~$m (\lone + m + 3)$.
	Note that in~$G \cup M'$ each middle node~$e^\text{mid}$ for the edge~$e=\{u,v\} \in E$ has distance~$(\lone + 1)$ to the two helper nodes~$h_v$ and~$h_u$ and distance more than~$2\lone$ to all other helper nodes and the demand nodes.
	Thus, in~$G \cup M$ the distance between any demand node and any middle node is more than~$\lone$.
	By the choice of~$\lone$, we have that~$(m+1)\lone > m(\lone + m + 3)$.
	Thus, each demand node~$d$ has a path of length strictly less than~$2\lone$ to the middle node~$e^\text{mid}$ with~$D(d,e^\text{mid}) = 1$.
	Hence, for each edge~$e = \{u,v\} \in E$ at least one of~$h_v$ and~$h_u$ needs to be matched to a demand node.
	Thus, $S$ is indeed a vertex cover in~$G$.\end{proof}

\section{Approximation Algorithm}
\label{sec: alg}

In this section, we discuss the main algorithm of this paper, which provably retains a constant factor approximation for a critical class of demand matrices. 
The main idea behind our algorithm is to benefit from the fact that shifting our view to design a higher degree network first and then transferring that result into matching opens up new possibilities.

We first define new terminologies that we need to describe our approximation algorithm, then go over the algorithm and at the end, we show the approximation factor of our algorithm.

\subsection{Preliminaries.}
We now go over terminologies that we need in the design of our algorithm.

\noindent \textbf{Demand graph.} 
The demand graph is a weighted graph. It is built by considering the demand matrix as the adjacency matrix of the graph. However, if two endpoints have already an edge in the infrastructure, we do not consider an additional edge in the demand graph. 

\noindent \textbf{Super-graph and super-node.}
The super-graph is an undirected simple graph, consisting of super-nodes. A super-node is a collection of nodes of an underlying graph. 
We consider the case that each super-node contains exactly $\alpha$ nodes of the infrastructure graph, for a fixed $1 < \alpha \le n$. \footnote{We point out that the fixed super-node size gives us the bounds that we are looking for, but acknowledge that one can consider a variable with different super-node sizes.}

Similarly, we can define the demand graph for the super-graph. We call that super-demand-graph. 


\begin{figure*}[t]
    \captionsetup[subfigure]{justification=centering}
        \centering
    \begin{subfigure}[b]{0.18\linewidth}
    \includegraphics[width=\linewidth, trim={25 15 30 25}, clip]{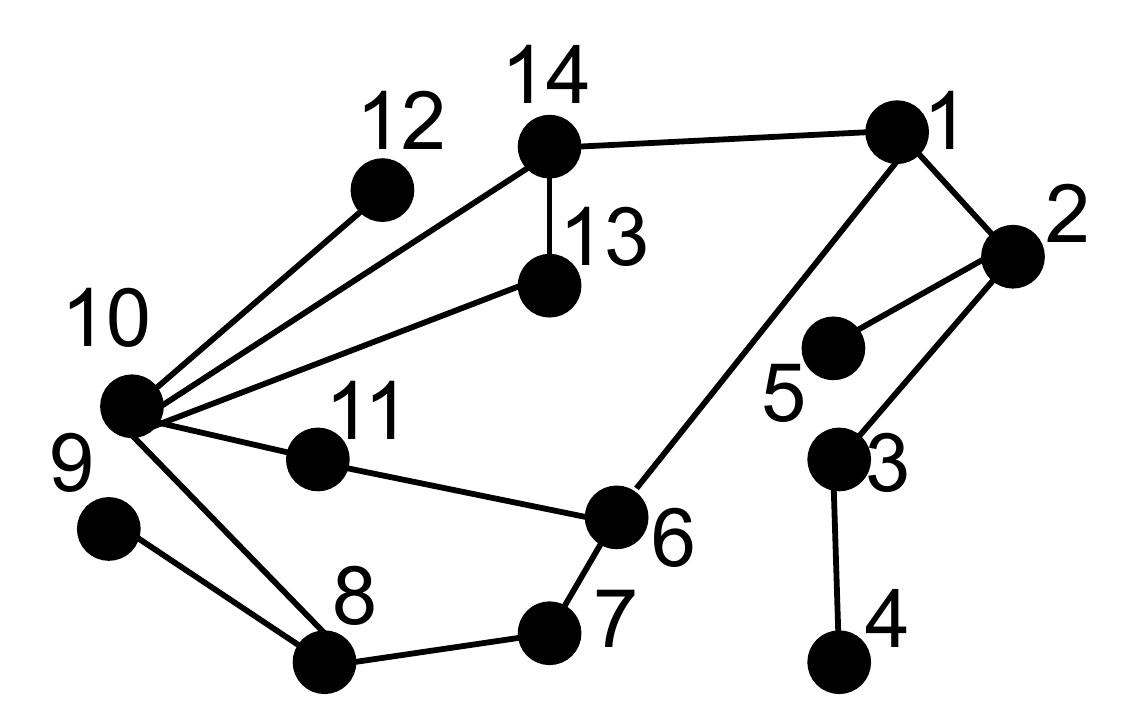}
    \caption{}
    \label{fig: main Graph}  
    \end{subfigure}
    \begin{subfigure}[b]{0.18\linewidth}
    \includegraphics[width=\linewidth, trim={25 15 30 25}, clip]{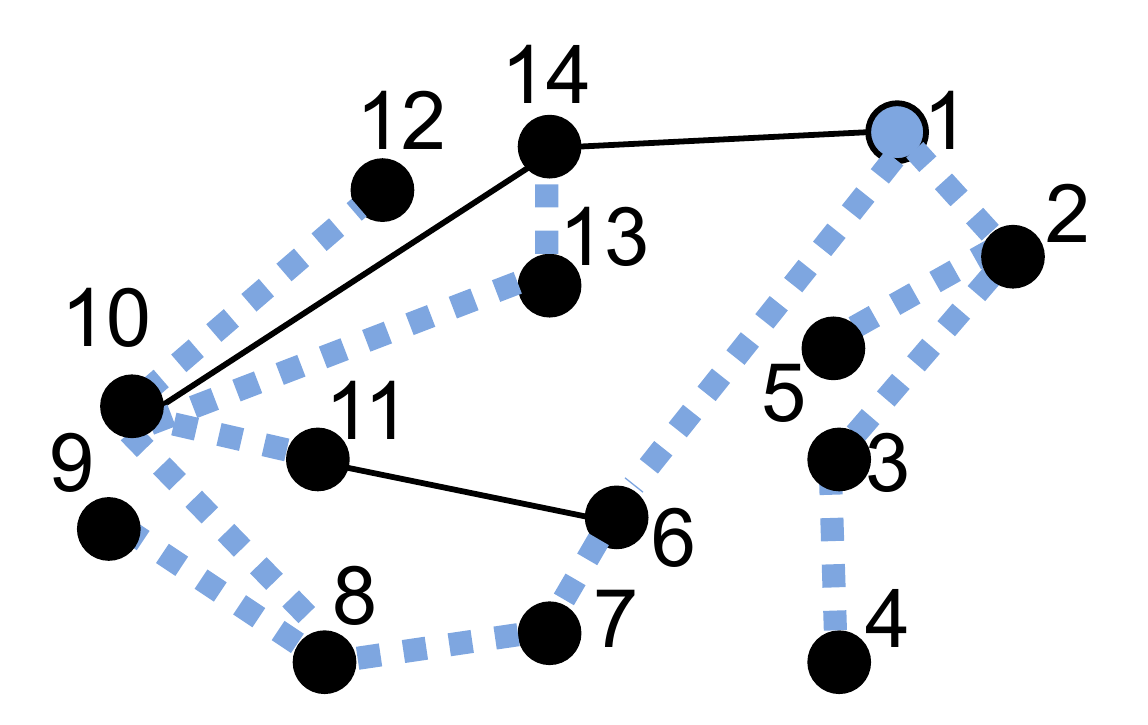}
    \caption{} 
    \label{fig: DFS}  
    \end{subfigure}
    \begin{subfigure}[b]{0.18\linewidth}
    \includegraphics[width=\linewidth, trim={25 15 30 25}, clip]{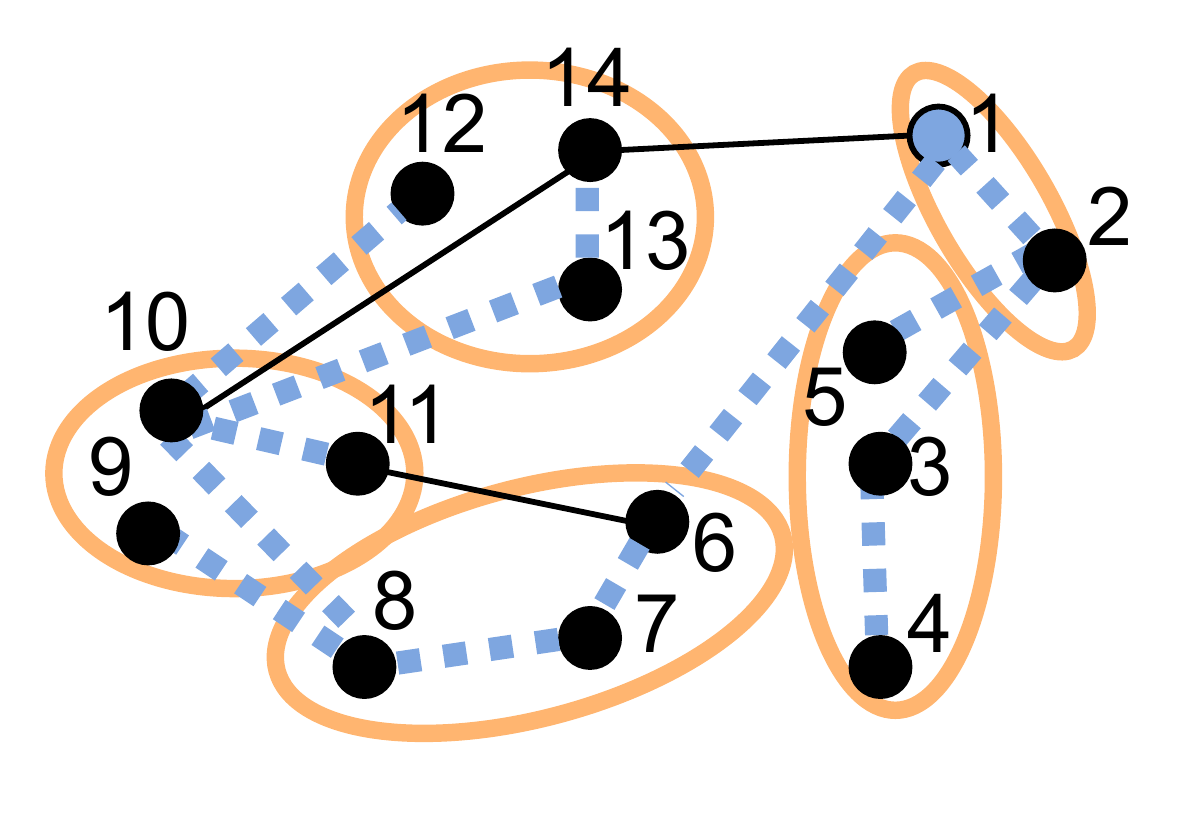}
    \caption{}
    \label{fig: positioning}  
    \end{subfigure}
    \centering
    \begin{subfigure}[b]{0.20\linewidth}
    \includegraphics[width=\linewidth, trim={35 20 35 20}, clip]{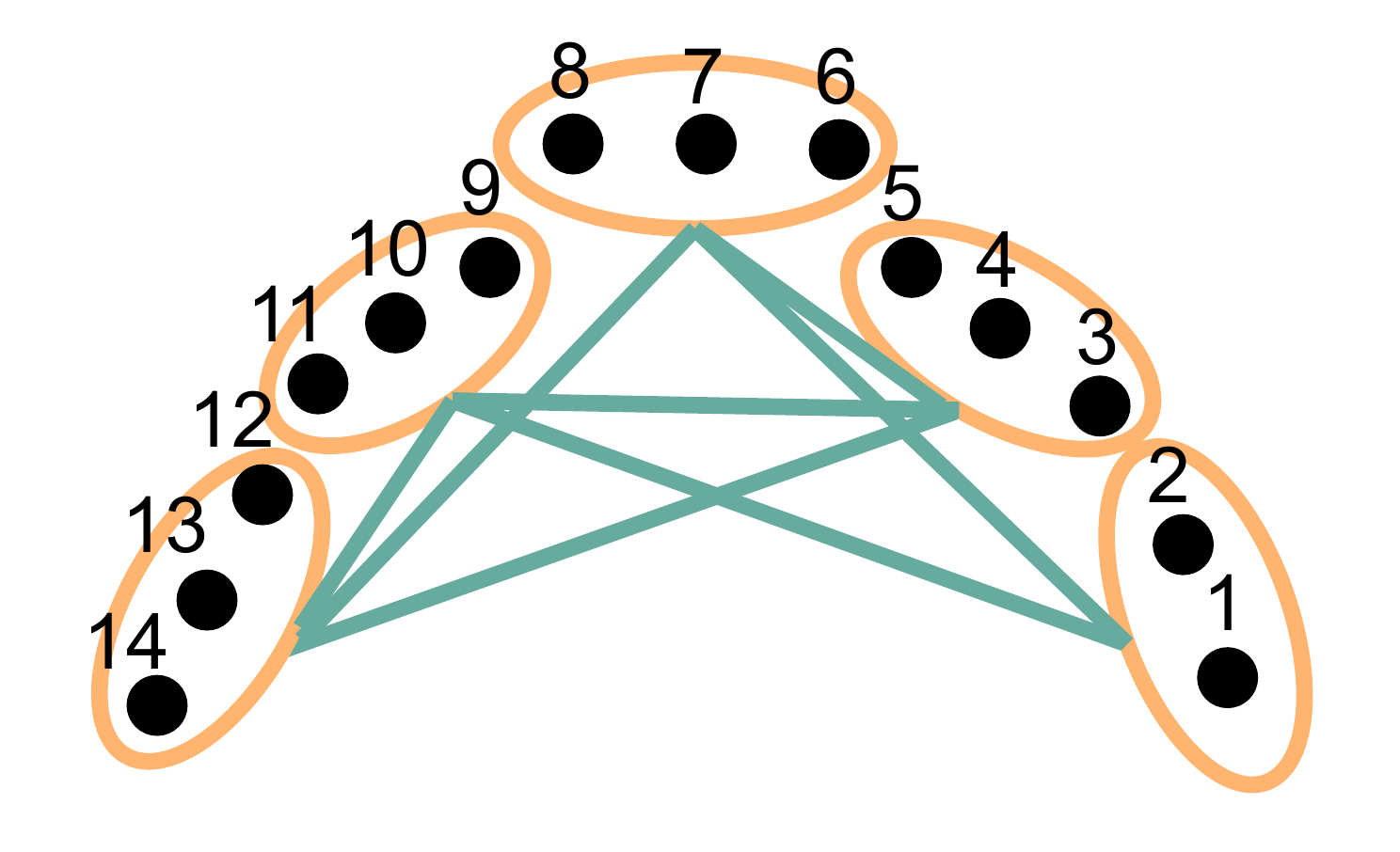}
    \caption{} 
    \label{fig: dan on super}  
    \end{subfigure}
    \centering
    \begin{subfigure}[b]{0.20\linewidth}
    \includegraphics[width=\linewidth, trim={0 0 0 0}, clip]{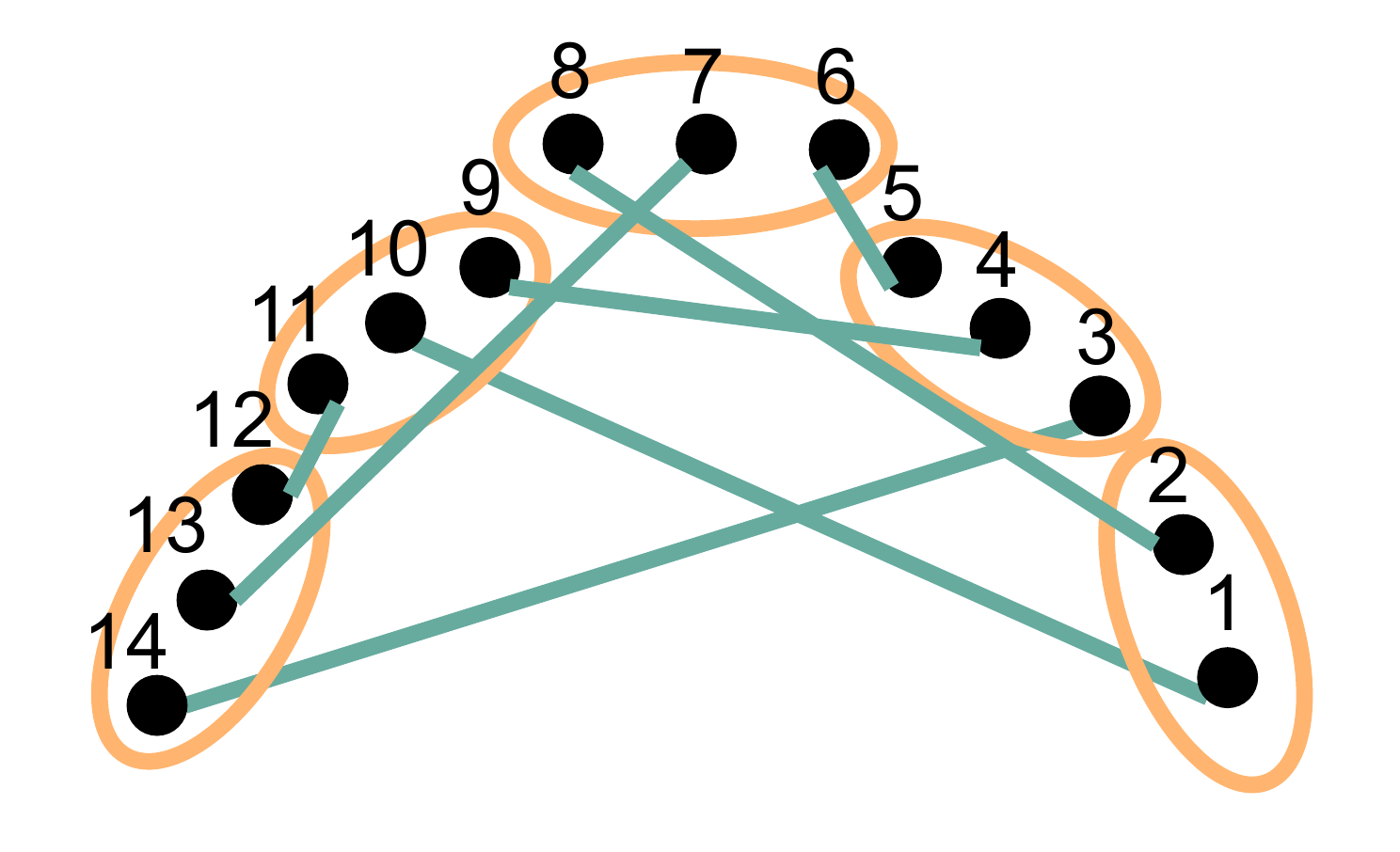}
    \caption{}
    \label{fig: inside super}  
    \end{subfigure}
    \caption{Steps of \superalgorithm considering a graph with $14$ nodes (Figure~\ref{fig: main Graph}), and super-nodes of size $\alpha = 3$. We first run a depth-first search (DFS) from the node in the upper-right, shown in blue (Figure~\ref{fig: DFS}). We then do a pre-order traversal of the tree from one of the deepest leaves, collecting nodes in groups of $3$(Figure~\ref{fig: positioning}, we consider the last group to have size only $2$). Figure~\ref{fig: dan on super} shows a DAN is built on top of the super-graph, and lastly in Figure~\ref{fig: inside super} we show how this DAN can be transformed back into matchings.
    }
    \label{fig: algorithm}
\end{figure*}
\subsection{Algorithm.}
Now we introduce our algorithm, \superalgorithm \footnote{Similar to how a spider creates its web from a combination of threads (matchings). DAN refers to Demand-Aware Network.}.
We start by discussing how super-nodes are formed, and then discuss how a demand-aware network (DAN) can be built on top of it. We conclude by discussing how the DAN on the super-graph can be transformed into matchings.

\noindent \textbf{Super-node creation.} 
From our infrastructure graph $G$ (that can be any connected graph, an example in Figure~\ref{fig: main Graph}) we create a \emph{super-graph} $S$ in two steps:
\begin{enumerate}
    \item Our algorithm first creates a spanning tree on the infrastructure graph (e.g. by running a \emph{Depth First Search}) from a node of $G$ (Figure~\ref{fig: DFS}).
    \item To form each super-node, we take the deepest node, and we consider the subtree of its grand-parent at distance $\alpha$.
    Then we select and delete $\alpha$ arbitrary deepest nodes from this subtree one by one, and add them to the super-node and then remove them from the sub-tree. We repeat this procedure until we group all nodes into super-nodes (Figure~\ref{fig: positioning}).
\end{enumerate}
Let us assume the output of the above steps as the node mapping $f_S$. For the super-graph $S$ the node set is $S(V) = f_S(V(G))$ and the edge set $E(S) = \{ \{f_S(u), f_S(v)\} \mid f_S(u) \neq f_S(v) , \{u,v\} \in E(G)\}$. For simplicity, in the rest of this section we assume~$n$ to be divisible by~$\alpha$ (later on, we will discuss how we overcome this assumption in practice, as our super-node creation algorithm can run without this assumption). 
The entries of the demand $D_S$ for $a, b \in V(S)$ are $$D_S(a,b) = \sum_{\substack{u, v \in V \\f_S(u) = a, f_S(v) = b}} D_{u,v}$$


\noindent \textbf{DAN on super-graph.}
After creating the super-graph, we use the construction of~\citet{avin2017demand} to build a demand-aware network (DAN)  on top of the super-nodes. We consider a degree at most $12 \cdot \avgdeg$ on top of the super-graph, where~$\avgdeg$ denotes the average degree of the super-demand-graph, that is the number of non-zeros per row/column in the matrix.

\noindent \textbf{Matching assignment inside a super-node.}
We now transform the edges of the DAN on top of super-nodes into a matching on the original graph. 
Thus, for each super-node~$v$ we can map each incident edge to a different node in~$f_S^{-1}(v)$ and obtain a matching in~$G$. See a visualization in Figure~\ref{fig: inside super}. In this transformation, we try to not add an edge that already exists in the infrastructure graph. 

\subsection{Analysis.}
In this section, we prove that \superalgorithm on very sparse demands and constant degree infrastructure graphs computes a constant factor approximation, and discuss its running time at the end. For analytical purposes, we restrict ourselves to demand graphs with an average degree of at most~$1/12$ (we discuss how this assumption can be improved later).
Note that the constructed DAN has a maximum degree of $12 \avgdeg \leq 12$, by assumption.\footnote{The proofs of statements marked by \appsymb{} are deferred to an appendix.}

\begin{lemma}{\appref{lem: good decomposition}}
    \label{lem: good decomposition}
    Any two nodes within a super-node have a distance of at most $2 \alpha$ in the infrastructure graph.
\end{lemma}
\appendixproof{lem: good decomposition}{
\begin{proof}
We prove by induction on the number of super-nodes created in the super-node creation algorithm. We prove a stronger statement than above: we prove that our algorithm additionally keeps the tree connected after removing already picked nodes.

When we have zero super-nodes, and we consider the deepest node in DFS, and all the nodes in the sub-tree of its grand-parent of distance $\alpha$. We know that we can go from any of the nodes in the subtree to any other node by traversing at most $2 \cdot \alpha$ edges (by simply going to the grand-parent from the source node and then to the destination node). Furthermore, we have at least $\alpha$ nodes in this subtree. Hence, a super-node can be created in this sub-tree, and the distance between any two nodes is at most $2 \cdot \alpha$. 
Furthermore, by iteratively grouping the leaves into the super-node, we can ensure that the tree remains connected.

Now let us assume that we have selected $k$ super-nodes, and consider nodes that are being selected in the super-node $k+1$. Given that after $k$ super-node creation the tree remains connected, we can simply use similar arguments to the zero case, to show that the distance between nodes in the selected super-node is at most $2 \cdot \alpha$ and the tree remains connected after grouping nodes into $k+1$ super-node. 
\end{proof}}

\noindent \textbf{Approximation ratio}. To compute the approximation ratio of \superalgorithm, we first require a few relations between adding a matching to a graph and demand-aware networks for the super-graph.

Lemma~\ref{thm: cost of opt} establishes a lower bound on the matching cost using demand aware networks for the demand $D_S$.

\begin{lemma}
\label{thm: cost of opt}{\appref{thm: cost of opt}}
Given a graph $G$ with demand $D$, and a super-graph $S$ with corresponding demand $D_S$ obtained by merging $\alpha$ nodes into super-nodes as defined by the node mapping $f_S$.
For any matching $M$ on $V(G)$, there exists a demand-aware network $H_S$ for $D_S$ with maximum degree at most $\alpha(1 + \Delta(G))$ such that
$\EPL_D(G+M) \geq \EPL_{D_S}(H_S)$.
\end{lemma}\appendixproof{thm: cost of opt}{
\begin{proof}
Given $H = G+M$ let $H_S$ be the graph obtained by merging nodes of $H$ into super-nodes in the same way that $S$ was obtained from $G$ using the node mapping $f_S$. The graph $H_S$ has maximum degree at most $\alpha(1 + \Delta(G))$.

Consider any $u,v \in V(G)$ with $u \neq v$ and let $P = v_0,\dots,v_\ell$ be a shortest path between $u$ and $v$ in $H$, where $v_0 = u, v_\ell =v$.
Let $P_S = f_S(v_0),\dots,f_S(v_\ell)$, which is a walk in $H_S$. Therefore $\dist_H(u,v) \geq \dist_{H_S}(f_S(u),f_S(v))$.
We now have that

\begin{align*}
& \EPL_D(H)  = \sum_{u,v \in V} \dist_H(u,v) \cdot D_{u,v}  \leq \\& \sum_{u,v \in V} \dist_{H_S}(f_S(u), f_S(v)) \cdot D_{u,v} = \EPL_{D_S}(H_S)
\end{align*}
\end{proof}
}

Given a demand-aware network for the super-graph with maximum degree at most $\alpha$, then a matching on the infrastructure graph can be found with cost only a constant higher than that of the demand-aware network.
\begin{lemma}
    \label{thm: supergraph to graph}
    Given a graph $G$ and demand $D$, and a super-graph $S$ with corresponding demand $D_S$ obtained by merging $\alpha$ nodes into super-nodes as defined by the node mapping $f_S$. For any demand-aware network $H_S$ with maximum degree at most $\alpha$ there exists a matching $M$ on $V(G)$ such that $\EPL_D(G+M) \leq 7\alpha \cdot \EPL_{D_S}(H_S)$.
\end{lemma}

\begin{proof}
    We construct a matching $M$ in the following way: initially $M=U=\emptyset$. For each edge $\{a,b\} \in E(H_S)$ pick any $u \in f^{-1}_S(a)$ and $v \in f^{-1}_S(b)$ with $u, v \notin U$, further add $\{u, v\}$ to $M$, and add $u,v$ to $U$. Since $\Delta(H_S) \leq \alpha$ and $|f^{-1}_S(c)| = \alpha$ for all $c \in V(H_S)$, this algorithm will always find a matching $M$ with $|M| = |E(H_S)|$.

    Now consider any $u, v \in V$ and let $P_S = a_0,\dots,a_\ell$ be a shortest path between $f_S(u)$ and $f_S(v)$ in $H_S$. For each edge $\{a_i, a_{i+1}\}$ of the path $P_S$ we have a corresponding matching edge $\{v^i_s, v^{i+1}_t\}$ in $M$. W.l.o.g. assume $f_S(v^i_s) = a_i$ and $f_S(v^{i+1}_t) = a_{i+1}$. We denote by $x-y$ a shortest path between $x$ and $y$ in $G$. We can find a walk from $u$ to $v$ of the following form: $P = u-v^0_s,v^1_t-v^1_s,v^2_t,\dots,v^{\ell-1}_s,v^{\ell}_t-v$. The ``$-$'' parts of $P$ are paths along edges of the infrastructure graph, between two nodes that are merged into the same super-node, which means these parts have length at most $2\alpha$. The ``,'' parts of $P$ are always edges from $M$. This means $P$ has length at most $(\ell+1)(2\alpha) + \ell$. This is at most $2\alpha$ for $\ell=0$ and at most $5\alpha\ell$ for $\ell \geq 1$.

    Let $H = G+M$. From the above, it follows that
    \begin{align*}
    \EPL_D(H) & = \sum_{u,v \in V} \dist_H(u,v) D_{u,v} \leq \\ & \sum_{u,v \in V} (2\alpha + 5\alpha \dist_{H_S}(f_S(u), f_S(v))) D_{u,v} \\ &= 2\alpha + 5\alpha \EPL_{D_S}(H_S) \leq 7\alpha \EPL_{D_S}(H_S)
    \end{align*}
    Where the last inequality holds, because $\EPL_{\ast}(\ast) \geq 1$ for any demand and any graph.
\end{proof}

An important property of demand-aware networks is that their cost can be lower bounded by a metric related to the demand matrix, namely conditional entropy, with a logarithmic factor in the maximum degree.
\begin{lemma}[DAN lowerbound \cite{avin2017demand}]
\label{thm: DAN lowerbound}
    For any demand $D$ and any demand-aware network $G$ with maximum degree at most $\Delta$ it holds that
    $\condentropy(D) / \log_2{(\Delta+1)} - 1 \leq \EPL_D(G)$, where $\condentropy(D)$ is the  conditional entropy of the demand matrix, that is $\condentropy(D) = \sum_{v \in V} d_v \cdot \sum_{u \in V} \log_2{\frac{d_v}{D_{u,v}}}$ with $d_v = \sum_{u \in V}D_{u,v}$.
\end{lemma}

We are now ready to prove the main statement of this section.
\begin{theorem}
\label{thm: constant factor superDAN}
    Given a graph $G$ with a demand graph $D$ of average degree at most $\frac{1}{\alpha}$, then for $\alpha = 12$ the \superalgorithm computes a matching $M$ for which
    $$ \EPL_D(G+M) \leq c \cdot \min_{\substack{\text{matching } M' \\ \text{on } V(G)}} \EPL_D(G+M') $$
    for some non-negative constant $c$ that depends only on $\alpha$ and $\Delta(G)$.
\end{theorem}
\begin{proof}
Let $S$ be the super-graph constructed from $G$ and $f_S$ the corresponding node to super-node mapping.
The super-demand $D_S$ has an average degree at most $1$.

Given $D_S$ as input, the algorithm by~\citet{avin2017demand} computes a demand-aware network $H_S$ with maximum degree at most 12, whose expected path length under $D_S$ is optimal up to a constant factor. More specifically, $\EPL_{D_S}(H_S) + 1 \leq c_1 \cdot \condentropy(D_S)$, for some non-negative constant $c_1$ that depends only on $\alpha$.
Furthermore, one can see that $\EPL_{D_S}(H_S) \leq 2 \cdot c_1 \cdot \condentropy(D_S)$, because $\EPL_{\ast}(\ast) \geq 1$ for any demand and any graph.
This in combination with Lemma~\ref{thm: DAN lowerbound} means it is a constant factor approximation for an $\alpha$-degree demand-aware network for $D_S$.

Using Lemma~\ref{thm: supergraph to graph} on $H_S$ we obtain a matching $M$ over $V(G)$ such that $\EPL_D(G+M) \leq 7\alpha \cdot \EPL_{D_S}(H_S) \leq 14 \cdot \alpha c_1 \cdot \condentropy(D_S)$. From the well-known decomposition (or grouping) property of entropy~\cite{infotheory2006,avin2017demand}, which intuitively means that merging two probability values into one by adding them, does not increase entropy, it follows that $\condentropy(D_S)) \leq \condentropy(D)$ and consequently we have $\EPL_D(G+M) \leq 14 \cdot \alpha c_1 \cdot \condentropy(D)$.

\newcommand{\Mopt}{M_\text{opt}}
From Lemma~\ref{thm: DAN lowerbound} and Lemma~\ref{thm: cost of opt} we know that the cost of an optimal matching $\Mopt{}$ is bounded from below by $\condentropy(D) / \log_2{\left(\alpha(1 + \Delta(G)) +1 \right)} - 1$.
By rearranging terms and setting $c_2 = 1/ \log_2{\left(\alpha(1 + \Delta(G)) +1\right)}$ we obtain $c_2 \cdot \condentropy(D) \leq \EPL_D(G+\Mopt{}) + 1$. Using again that $\EPL_{\ast}(\ast) \geq 1$ for any demand and any graph, we obtain $\EPL_D(G+\Mopt{}) \geq \frac12 c_2 \condentropy(D))$, which proves that the matching $M$ computed by \superalgorithm is a constant factor approximation (with factor $c = 28 \cdot \alpha \cdot c_1/c_2$).
\end{proof}

We remark that Theorem~\ref{thm: constant factor superDAN} would be able to achieve $\alpha=5$ if we replace the algorithm from~\citet{avin2017demand} by the recent algorithm from~\citet{figiel2023demand}; the proof is analogous. However, we point out that the result of this recent paper is not needed to ensure constant approximation of our algorithm.

\noindent \textbf{Running time.} The running time of our algorithm is $O(n^2 \cdot \log  n)$. 
Grouping nodes into super-nodes using a depth-first search algorithm depends on the number of edges of the infrastructure graph which is at most $O(n^2)$.

Demand-graph requires a running time of $O(n^2)$, as it requires going over the whole demand matrix.
Then, running the DAN algorithm~\cite{avin2017demand} has a running time of $O(n^2 \cdot \log  n)$.
Turning edges of DAN to matchings also takes $O(n)$, as turning $\alpha$ edges of a super node to matchings takes a constant time. Hence, in total, the running time of the algorithm is $O(n^2 \cdot \log  n)$\footnote{The running time of the algorithm can be improved to $O(n \cdot \log n)$, if the sparsity of demand matrix is known beforehand and the demand matrix is given in a sparse representation.}.



\section{Heuristics}
\label{sec: heuristic}

We will experimentally compare \superalgorithm against several natural heuristics that we discuss below.
Note that most of these heuristics do not provide any theoretical guarantees on the solution quality.
%

\noindent \textbf{Greedy.} 
This algorithm matches pairs of nodes one by one. 
A pair of nodes is \emph{valid} if it does not increase the degree of nodes to higher than $1$.
The \algoname{Greedy} algorithm starts by sorting all pairs of nodes based on their demand in descending order. 
Starting with the edge of highest demand, it considers the next valid edge in the sorted list as the next matching edge. 
The running time of this algorithm is $O(n^2 \cdot \log n)$, which comes from the sorting of the list.
Among all heuristics, this is the easiest algorithm to implement, given that it does not rely on other algorithms. 

\noindent \textbf{Matching on demand.}
This heuristic builds on top of the well-known maximum weighted matching~\cite{Matching1,Matching2}. 
The \emph{Matching on demands} algorithm uses the \emph{demand graph} that we introduced before, in which each edge is weighted by its demand, except the infrastructure edges. The algorithm then considers the maximum matching on the demand graph as its output. The running time of this algorithm is $O(n^3)$, given the currently best maximum weighted matching algorithm~\cite{DuanP14}. 

\noindent \textbf{SuperChord.}
Based on the idea of creating super-graphs, we 
    aim
    to benefit from the well-known Chord~\cite{StoicaMKKB01} protocol. 
    To this end, we create a super-graph by combining $x = \frac{W(n \cdot \ln 2)}{\ln 2}$  consecutive nodes, where $W$ represents Lambert $W$ function, and $\lg$ the natural logarithm function\footnote{If the number of super-nodes is not a power of two, we consider the super-graph with closest and smaller power of two as the number of super-nodes.}. 
    We then build the Chord on the $\frac{n}{x}$ super-nodes of the super-graph. The resulting graph has degree $\log(\frac{n}{x})$. 
    We set $x = \frac{W(n \cdot \ln 2)}{\ln 2}$ as it ensures $x = \log(\frac{n}{x})$. 
    This allows us to view the $\log(\frac{n}{x})$ outgoing edges of a super-node as matching edges initiated from the nodes within the super-node.
    Observe that the super-graph can be created in linear time in $n$, and adding $x$ edges on each of $\frac{n}{x}$ super-nodes only takes $O(n)$ time.
    Thus, with a running time of $\Theta(n)$, the algorithm is relatively fast\footnote{We remark that computing the cost of this algorithm depends on the size of demand graph.}.
    We note that the Chord algorithm ensures a $\log n$ diameter (considering degree $\log n$)~\cite{StoicaMKKB01}.
    As the size of super-nodes is $O(\log n)$, we can ensure that any two nodes can reach each other via a shortest path of length $O(\log n)$, which in turn implies that \algoname{SuperChord} is an~$O(\log n)$ approximation.


\section{Experimental Evaluation}
\label{sec: exp evaluation}
In this section, we evaluate the weighted average shortest path length and the running time of our approximation algorithm, the heuristics.
\begin{enumerate}[label=\textbf{Q\arabic*.}]
    \item How fast are our algorithms in practice?

    \item How do our algorithms perform on real-world demands?
    \item Under which demand parameters do our algorithms perform better?
    \item What is the effect of the underlying infrastructure graph on the performance of algorithms?
\end{enumerate}
We believe answering the above questions would help developers in selecting the right algorithm for their use cases.


\subsection{Demand matrices.}
To evaluate our results, we consider both \emph{real-world} instances and \emph{synthetically} generated ones. By doing so, we first show the benefits of each algorithm in the wild, and then suggest the best algorithmic choices for possible use cases in the future. 

\noindent \textbf{Zipf distribution demand.} The Zipf distribution~\cite{Zipf2} has shown to be an effective estimator for traffic frequency distribution in data center networks~\cite{zipf1929relative,tracecomplexity}.
The Zipf distribution depends on a parameter $\zeta > 0$, which indicates the skewness of the distribution. With lower values of $\zeta$, the distribution is more skewed. We use a range $\zeta \in [2,10]$ for Zipf values in our evaluations.
Concretely, for given $\zeta$ and $n$ elements, the probability mass function for an $x \in [1,n]$ is determined by $f(x) = \left({x^{\zeta} \cdot \left(\sum_{i=1}^n \frac{1}{i^{\zeta}}\right)}\right)^{-1}$. 
When running our experiments for various values of~$\zeta$, we normalize the sum to ensure the total demand remains the same.  

\noindent \textbf{Sparse demand.} 
In order to test our algorithms on an even wider range of possibilities, we generate random demand matrices with controlled sparsity. For a sparsity parameter $\gamma$, we ensure that each cell of a matrix has high demand (determined by the user, we considered $100$) with probability $1-\gamma$. We consider the range $\gamma \in [0.1,0.9]$ for sparsity values.


\noindent \textbf{Real-world demands.}
Our code has been tested on a range of real-world data center traces~\cite{tracecomplexity} that has been the base of comparison for many previous works~\cite{hanauer2022fast,SeedTree,PowerOfMatching}.
In particular, we focused on the Meta (formerly known as Facebook) dataset~\cite{facebook}. This dataset contains communication between racks and servers within three data center clusters (Database, Web Services, and Hadoop, sorted by their number of nodes) which we call $A$ to~$C$ respectively. We focus on the communications between racks. We summarize each dataset into a list, in which the frequency of communication between each rack pair is listed.

Furthermore, we consider a set of $66$ instances from SuiteSparse matrix collection (formerly known as the University of Florida Sparse Matrix Collection)~\cite{DavisH11}, covering various applications. We have chosen symmetric matrices with up to $10,000$ rows, that also have positive values.

\subsection{Infrastructure graphs.}
In our evaluations, we have used symmetric infrastructure graphs. We believe that the following infrastructure graphs can give us the insight that we need to incorporate our algorithms in real-world applications.

\noindent \textbf{Ring.}
A ring is the simplest symmetric infrastructure that ensures the connectivity of the graph. Hence, this structure echoes the effect of the added matching the most. 
The ring graph has been a backbone of fundamental advancements in the design of reconfigurable networks~\cite{StoicaMKKB01, MirrokniTZ18}.
We believe the ring is a good candidate to be the default infrastructure to run our experiments on: it has a high diameter, symmetric, and simple connected graph, and hence can show the effect of various algorithms in cost reduction more clearly. We mention explicitly when using other infrastructure graphs in our experiments.

\noindent \textbf{2D and 3D Torus.}
A torus is a natural extension of a ring, which ensures more connectivity in the infrastructure graph. 
A torus is a grid that preserves symmetry by connecting border nodes to each other.
The grid-like structures have been the basis of previous studies on the effect of adding a matching, for example when discussing Kleinberg's model for small world networks~\cite{kleinberg2000navigation,easley2010networks}. 
In our evaluations, we considered both 2D torus structures (where each node has 4 neighbors) and 3D torus structures (where each node has 8 neighboring nodes).

\subsection{Results.}
Our code is written in python~$3.10$, benefiting from networkx~\cite{NetworkX} and gurobipy~\cite{gurobi} libraries. Our visualizations use Matplotlib~\cite{Matplotlib}. The code was executed on a machine with Intel\textsuperscript{\textregistered} Xeon\textsuperscript{\textregistered} CPU E5-1620 CPU with a clock frequency of 3.60GHz, and 64GB RAM. 

We now focus on answering questions proposed at the beginning of the section, showing how our algorithms behave given the above-mentioned demand matrices and infrastructure graphs. Before that, we want to point out that we also tested an MIP-formulation (detailed in Appendix~\ref{sec: MIP}) solved with Gurobi optimizer~\cite{gurobi}.
We do not include it in our plots as it hit our time limit of $1$ hour per instance already on instances with $\approx20$ vertices. 
Notably, even the LP-relaxation could not solve the instances with $\approx 200$ vertices within the time limit.\footnote{We tried the LP-relaxation for lower bounds. However, we do not discuss these bounds as we only can solve a few smaller instances with the LP and the LP-bound was on average a factor of $\approx 2$ away from the MIP-solution (when we have both).}

In the \superalgorithm, when we transform edges of the super-graph to matching, i.e., picking nodes inside corresponding super-nodes to connect to each other, we select the pair of nodes that have highest demand between them.
Furthermore, we noticed in our experiments that for algorithms that use super nodes, sometimes these supernodes do not cover all nodes in the infrastructure graph (i.e., the size of the infrastructure graph is not divisible by the size of a supernode). 
In such a case, we run \algoname{Matching on demand} on those remaining nodes to complete the matching.

\begin{figure}[t]
    \centering
    \includegraphics[width=0.7\linewidth, clip, ,trim={5 5 5 0}]{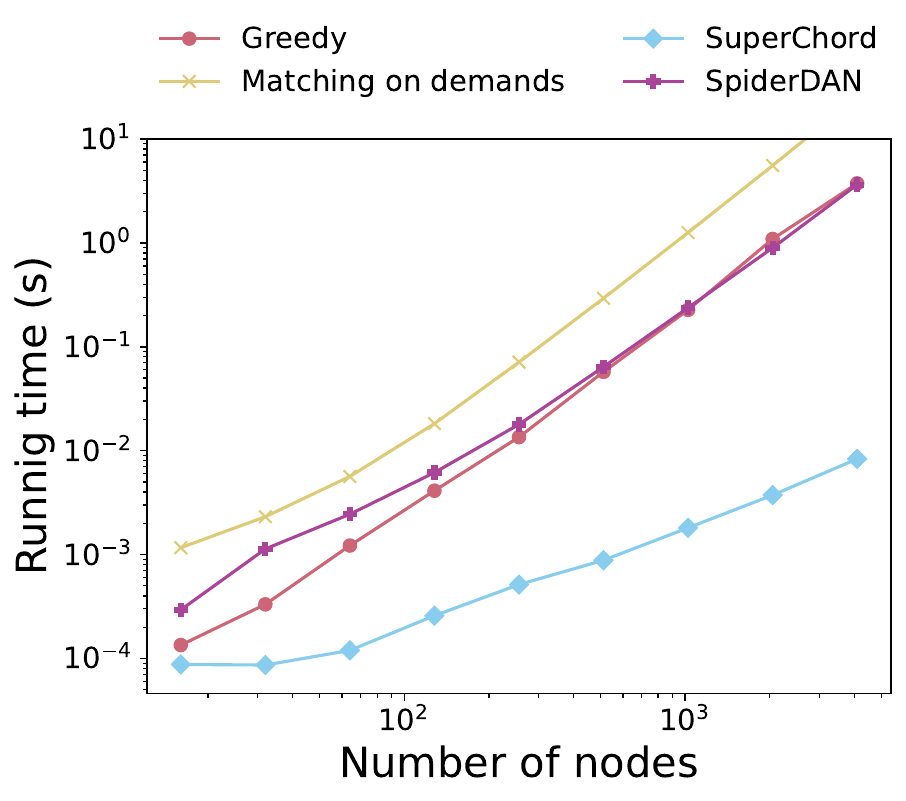}
    \caption{ 
    The running time of all of our algorithms is displayed.
    The number of vertices are powers of two (up to 4096 vertices). 
    We have considered randomly generated demand matrices with sparsity value $0.9$. 
    We capped off the running times at $10$ seconds, and showed it in log-log format for better visibility.
    }
    \label{fig: runtime}
\end{figure}

\begin{figure*}[t]
    \captionsetup[subfigure]{justification=centering}
    \centering
    \begin{subfigure}[b]{0.35\textwidth}
    \centering
    \includegraphics[width=\textwidth, clip,,trim={5 5 5 0}]{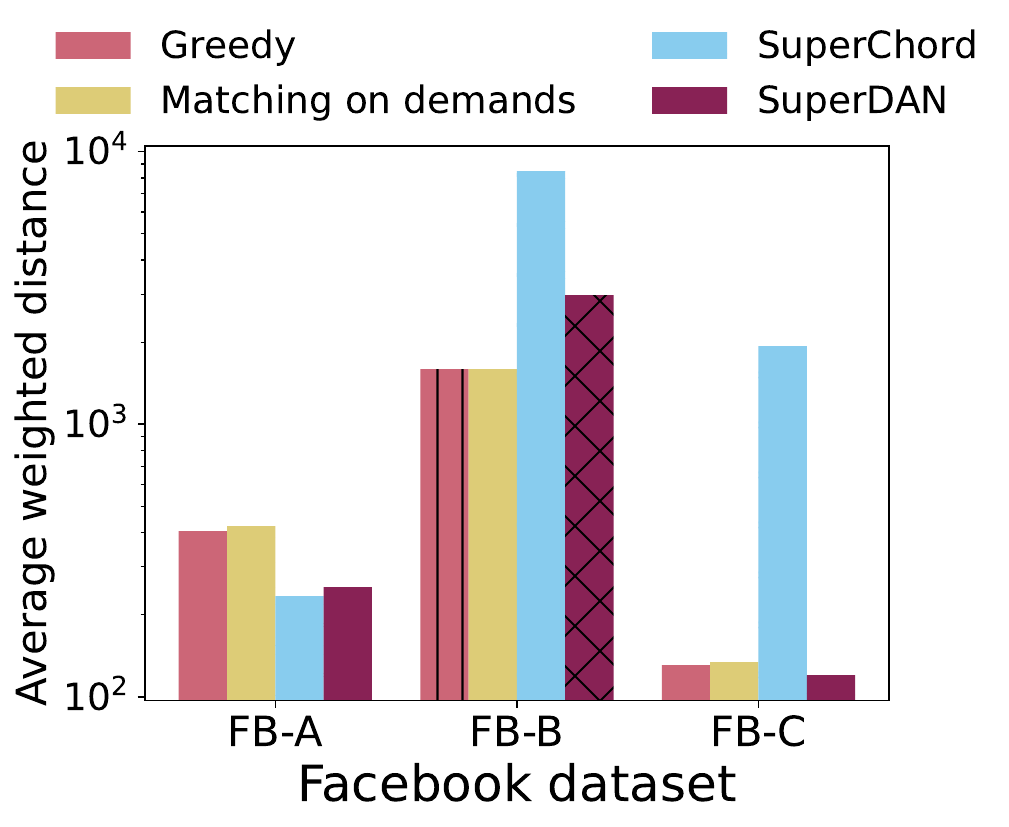}
    \caption{Data sets from Facebook.}
    \label{fig: real-new}  
    \end{subfigure}
    \begin{subfigure}[b]{0.64\textwidth} \includegraphics[width=\linewidth, clip]{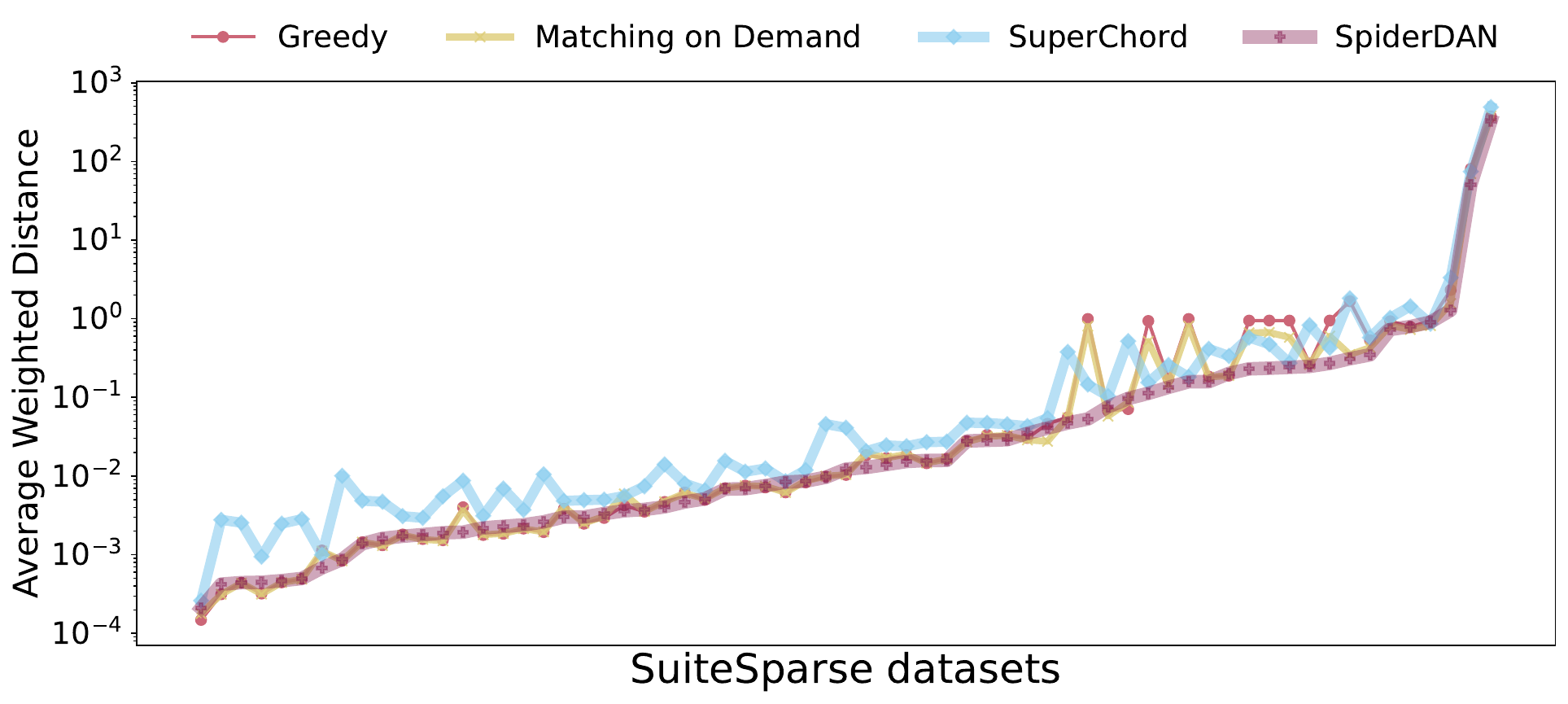}
    \caption{Data sets from SuiteSparse matrix collection.}
    \label{fig: Suite}  
    \end{subfigure}

    \caption{Left: the results on the three datasets from Facebook. Right: Results for $66$ instances from SuiteSparse matrix collection. The instances are sorted by the quality of \superalgorithm{}. }
    \label{fig: demand}
\end{figure*}

\noindent \textbf{A1. Running time of algorithms.} 
All our provided algorithms except \algoname{Matching on demands} are sufficiently fast on the graph sizes that we expect in practice (with up to a couple of thousands of nodes) with running times measured in seconds, see \Cref{fig: runtime} for a plot.
There are, however, notable differences between the algorithms.
\algoname{SuperChord} is by far the fastest algorithm as it essentially ignores the demands (which can be $O(n^2)$ many).
Next are \algoname{Greedy} and \superalgorithm{} that show very similar running times.
\algoname{Matching on demands} is clearly the slowest algorithm with the computation of a maximum weight perfect matching being the overall most costly operation.


\noindent \textbf{A2. Results on real-world data sets.} 
Here we focus on the Facebook and SparseSuite datasets. 
As shown in Figure~\ref{fig: real-new}, the results are mixed on the Facebook dataset. 
\algoname{Matching on demands} and \algoname{Greedy} perform very similar. 
\superalgorithm{} performs very well on clusters A and C, but is a bit worse on cluster B.
Previous work~\cite{facebook} has shown that cluster B has a higher average demand degree, which is the reason behind the slightly higher cost across different algorithms.
Moreover, it aligns with our theoretical findings (see Theorem~\ref{thm: constant factor superDAN}), that \superalgorithm{} performs better on low average demand degree.

Given that \algoname{SuperChord} performs poorly on clusters B and C, we suspect the demands to be very skewed in these instances. 
In contrast, \algoname{SuperChord} gives the best results on cluster A.

As shown in Figure~\ref{fig: Suite}, on sparse instances, our algorithms (except the demand-agnostic \algoname{SuperChord}) perform more or less the same, but in most of instances where there is a noticeable difference, \superalgorithm{} seems to perform better.

\noindent \textbf{A3. Effect of demand parameters.}
Here, we discuss the effects of the parameters of three synthetic demands, namely the sparsity value ($\gamma$), and the zeta value of the Zipf distribution ($\zeta$). We go through the details of the results for each parameter.

\begin{figure*}[t]
    \captionsetup[subfigure]{justification=centering}
    \centering
    \begin{subfigure}[b]{\textwidth} 
    \centering
    \includegraphics[width=0.7\linewidth, clip,trim={0 14 5 5}]{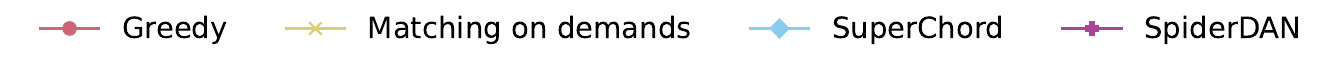}
    \label{fig: legendTwo}  
    \end{subfigure}
    \begin{subfigure}[b]{0.32\textwidth} \includegraphics[width=\linewidth, clip]{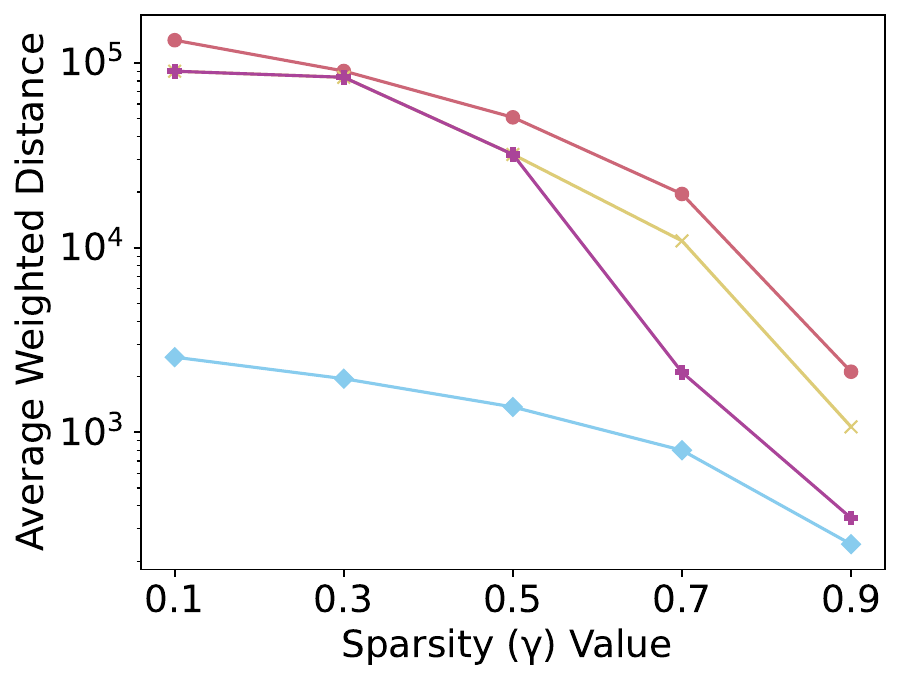}
    \caption{Varying sparsity.}
    \label{fig: sparse}  
    \end{subfigure}
    \begin{subfigure}[b]{0.32\textwidth}
    \centering
     \includegraphics[width=\textwidth, clip]{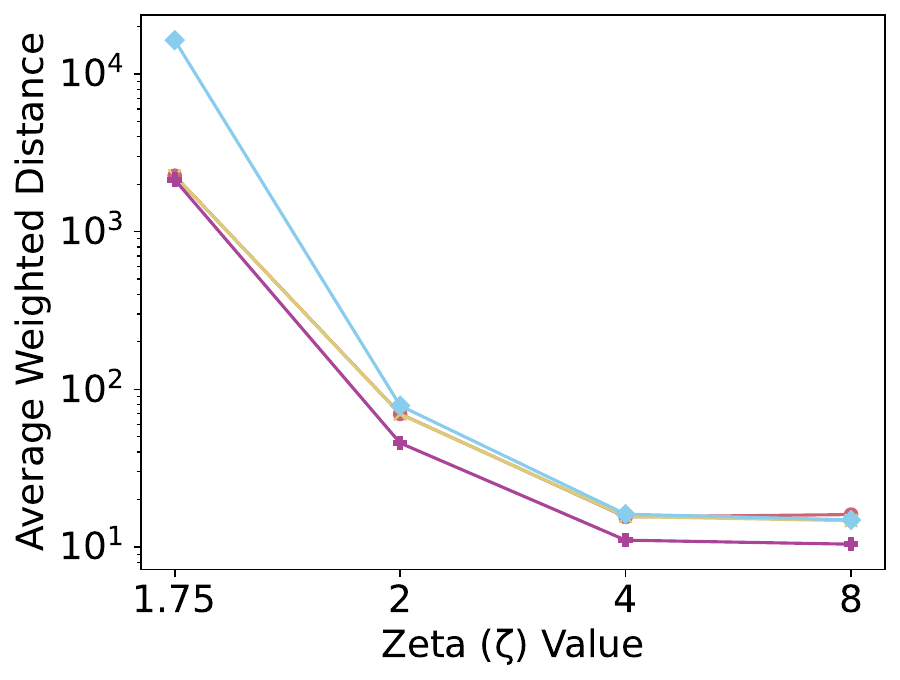}
    \caption{Varying zeta values.}
    \label{fig: zipf}  
    \end{subfigure}
    \begin{subfigure}[b]{0.32\textwidth}
    \centering
    \includegraphics[width=\textwidth, clip,trim={5 5 5 0}]{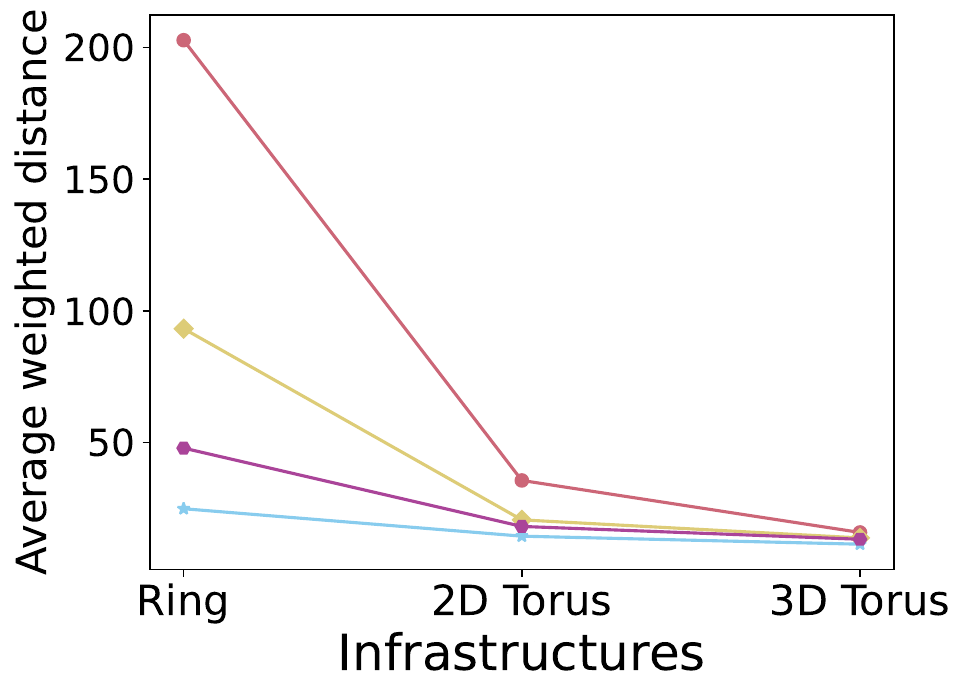}
    \caption{Various infrastructures.}
    \label{fig: infra}  
    \end{subfigure}
    \caption{Effect of various parameters on the approximation ratio of algorithms (the cost of the respective algorithm divided by the cost of ring).
    Figure~\ref{fig: infra} considers on $4096$ node (as it is even and a power of six, so we can have both 2D and 3D Torus with this size), and  we used sparse instances with $\gamma =0.9$. The two other figures are based on $4096$ and $1024$ nodes, due to the time limit that we set for each instance.
    }
    \label{fig: params}
\end{figure*}

\begin{itemize}
    \item \textbf{Sparsity of the random distribution.} 
    As expected, with more sparse demand our algorithms provide solutions of lower cost, see \Cref{fig: sparse}.
    A sparser demand can cause a higher fraction of the demand being directly covered by the edges in the matching, that is, the demand pairs will have a distance of one, and hence a lower cost.
    On the other hand, the demand-agnostic \algoname{SuperChord} performs very well when faced with less sparse demand, especially in this case when the non-zero demands are equal.
    The random demands are quite uniform, which benefits \algoname{SuperChord}.
    \item \textbf{Zeta of the Zipf distribution.} 
    Similar to before, based on our results shown in \Cref{fig: zipf}, we can see that as the zeta of Zipf distribution grows, i.e. data becomes more uniform, the algorithms perform better. 
    However, in contrast to the previous case, here, the values of non-zero demands can be different, therefore \algoname{SuperChord} does not have the edge that it had beforehand, and \superalgorithm performs better than it and other algorithms. 
\end{itemize}

\noindent \textbf{A4. Effect of infrastructure graphs.}
Given the selected set of infrastructures, we can observe how an increase in the average degree of infrastructure graphs affects the cost of our algorithms. 
In particular, we believe our algorithms echo the inherent improved average distances in the infrastructure, as we saw a rapid improvement going from a ring (essentially a 1D torus) to a 2D torus. 
However, the improvement is much less when going from 2D to 3D torus, as can be seen in Figure~\ref{fig: infra}. 
Moreover, it can be seen that the ranking of the algorithms by solution quality stays the same for 1D, 2D, and 3D torus.
Hence, only considering the ring in our other experiments highlights the differences between the algorithms.


\subsection{Summary \& Outlook}

We first start by recapping the answers to the proposed questions:
\begin{enumerate}[label=\textbf{A\arabic*.}]
    \item With \algoname{SuperChord} being the by far fastest algorithm, \superalgorithm{} and \algoname{Greedy} are on shared second place and are still fast enough on the large real-world instances.
    \item Our results indicate that our algorithms using super-graphs, in particular \superalgorithm, can be a good option to reduce the cost in real-world instances, in conjunction with other greedy algorithms. 
    \item We observed that our algorithms can exploit the underlying demand structure, to provide close to optimal outcomes. In particular, part of our algorithms show promising results with high sparsity and high zeta values.
    \item We observed that our algorithm can utilize the underlying infrastructure graph to enhance their outcome, echoing the reduction in the average distance of the infrastructure graphs.
\end{enumerate}
In summary, we can recommend \algoname{SuperChord} for uniform demands: it provides the best and fastest solutions in this case.
However, for skewed demands especially in real-world datasets, the heuristic can perform quite poorly.

For each heuristic there is a real-world dataset where \superalgorithm provides solutions of better quality than the heuristic.
Overall \superalgorithm provides comparable results to the heuristics; there is no instance where it is outperformed significantly.
Given that it comes with some guarantees on the solution quality, we recommend it when solving real-world instances.

\section{Conclusion}
\label{sec: conclusion}

In this paper, we tackled the problem of minimizing the demand-aware average shortest path, via matching addition. 
Our goal is to augment the given physical network based on communication frequencies. 
We provided insights into its computational complexity and exact and efficient algorithms.
We started exploring the computational complexity of \problem, showing NP-hardness even in restricted cases and providing the constant-factor approximation algorithm \superalgorithm for highly skewed sparse demand matrices. 
To argue about general demand matrices we performed an extensive empirical evaluation. We thereby compared the \superalgorithm algorithm together with various heuristics to the exact solution (provided by a mixed integer program) on a series of real-world and synthetic datasets. 

Our paper opens interesting directions for future work. Observe that our \superalgorithm is designed for low average degrees in the demand graph. The lower bound, on the other hand, is based on conditional entropy and is constant for low-entropy demand matrices. An open question is whether it is possible to find a constant approximation for certain low entropy demand matrices that are not covered by our method. One particularly engaging example is the ring demand graph where the infrastructure graph is a different ring. 
As a contribution to the research community, and to ensure reproducibility, we open-source our code and experimental artifacts at the following URL:

\noindent\url{https://github.com/inet-tub/SuperDAN}



\bibliographystyle{plainnat}
\bibliography{main}

\clearpage
\appendix
\section{Omitted Proofs}
\label{app: omitted proofs}
In this section, we overview the proofs omitted in the main body of the paper.
\appendixProofs

\begin{table}[t]
\caption{A summary of variables used in the MIP.}
\begin{center}
\small
\begin{tabularx}{\linewidth}{cX}
   \hline 
   Input variables & 
   \cellcolor{gray!20} Description
    \\ \hline
    \cellcolor{gray!20} $D_{u,v}$ & Demand between nodes $u$ and $v$.
    \\ \hline
    \cellcolor{gray!20} $\deg_{u}$ & Degree of node $u$.
    \\ \hline
    \hline
    MIP variables & 
   \cellcolor{gray!20} Description
    \\ \hline
    \cellcolor{gray!20}  $a_{u,v}$ & Indicating whether the edge $(u,v)$ is added to the matching or not 
    \\ \hline
    \cellcolor{gray!20}  $\dis(u,v)$ & Distance between nodes $u$ and $v$.
    \\ \hline
    \cellcolor{gray!20}  $y^{u,v}_{w} $ & Binary variable indicating if shortest path between $u$ and $v$ goes through $w$ or not. 
    \\ \hline
\end{tabularx}
\end{center}
\label{table: parameters}
\end{table}
\begin{algorithm}
\caption{Mixed Integer Program to Compute Optimal Solution}
\label{alg: Mixed}
\begin{algorithmic}[1]
\State \textbf{Minimize $\sum_{u,v \in V} \dis(u,v) \cdot D_{u,v}$} \label{line: opt}
\State $\dis_{u,v} = 1 \hfill  \forall (u,v) \in E$  \label{line: edge in inf}
\algdef{SE}[FOR]{For}{EndFor}[1]%
  {\algorithmicfor\ #1\ \algorithmicdo}%
  {\algorithmicend\ \algorithmicfor}%
\algtext*{EndFor} 

\For{$u,v \in V \ \& \ u\neq v \ \& \ (u,v) \notin E$}
\label{line: for edge}
\State $y^{u,v}_{w} \in \{0,1\}$ 
\State $a_{u,v} \in \{0,1\}$ 
\State $a_{u,v} = a_{v,u}$ 
\State $\dis(u,v) \ge 1 $  \label{line: dis act 1}
\State $\dis(u,v) \le a_{u,v} + (1-a_{u,v}) \cdot M$\label{line: dis act 2}
\State $\dis(u,v) \le \dis(u,w) + \dis(w,v)  \hfill  \forall w \in V$ \label{line: dis all 1}
\State $\dis(u,v) \ge \dis(u,w) + \dis(w,v) + (y^{u,v}_{w}-1) \cdot M\;  \hfill  \forall w \in V$ \label{line: dis all 2}
\State $\sum_{w \in V \setminus \{u,v\}} y^{u,v}_{w} + a_{u,v} = 1$ \label{line: dis all 3}
\EndFor
\For{$u \in V$}
\State$\sum_{\forall v \in V \ \& \ v \neq u \in V \ \& \ (u,v) \notin E} a_{u,v} = 1 \hfill$  \label{line: active edge}
\State $dis_{u,u} = 0 \hfill $ \label{line: zero dis}
\EndFor
\end{algorithmic}
\end{algorithm}

\section{Mixed Integer Program}
\label{sec: MIP}
In this section, we detail our Mixed Integer Program (MIP), Program~\ref{alg: Mixed}, as an exact solution to the problem. A summary of variables used in our MIP is in Table~\ref{table: parameters}.

Our goal for this MIP is to minimize distance times the demand for each pair of nodes (Line~\ref{line: opt}). In doing so, we respect the edges of the infrastructure graph by setting the distance between their two endpoints equal to one (Line~\ref{line: edge in inf}) and the distance of a node to itself is zero (Line~\ref{line: zero dis}).
Our MIP uses a symmetric binary variable $a_{u,v}$ to decide whether it wants to add a matching edge between nodes $u$ and $v$. It therefore goes over all pairs of nodes $u \neq v$ that do not already have an edge in the infrastructure graph in Line~\ref{line: for edge}.

In order to have a matching, each node should have one active edge (Line~\ref{line: active edge}).
Lines~\ref{line: dis act 1} and~\ref{line: dis act 2} ensure that an active edge has a distance of one. We then force shortest path distances between all other nodes in Lines~\ref{line: dis all 1} to~\ref{line: dis all 3}.

\end{document}